\documentclass[12pt,preprint]{aastex}
\begin{document}
\title{SDSSJ102111.02+491330.4: A Newly Discovered Gravitationally Lensed Quasar\thanks{Observations reported here were obtained at the MMT Observatory, a joint facility of the University of Arizona and the Smithsonian Institution.}}
\author{
Bart~Pindor,\altaffilmark{2,3}
Daniel~J.~Eisenstein,\altaffilmark{4}
Michael~D.Gregg,\altaffilmark{5,6}
Robert H. Becker,\altaffilmark{5,6}
Naohisa Inada,\altaffilmark{7}
Masamune Oguri,\altaffilmark{8,9}
Patrick~B. Hall,\altaffilmark{10}
David E. Johnston,\altaffilmark{9}
Gordon T. Richards,\altaffilmark{9}
Donald~P.~Schneider,\altaffilmark{11}
Edwin~L.~Turner,\altaffilmark{9}
Guido~Brasi,\altaffilmark{4}
Philip~M.~Hinz,\altaffilmark{4}
Matthew~A.~Kenworthy,\altaffilmark{4}
Doug~Miller,\altaffilmark{4}
J.~C.~Barentine,\altaffilmark{12}
Howard~J.~Brewington,\altaffilmark{12}
J. Brinkmann, \altaffilmark{12}
Michael~Harvanek,\altaffilmark{12}
S.~J.~Kleinman, \altaffilmark{12}
Jurek~Krzesinski,\altaffilmark{12,13}
Dan~Long,\altaffilmark{12}
Eric~H.~Neilsen,~Jr.,\altaffilmark{14}
Peter~R.~Newman,\altaffilmark{12}
Atsuko~Nitta,\altaffilmark{12}
Stephanie~A.~Snedden,\altaffilmark{12}
and Donald~G.~York\altaffilmark{15}
}

\altaffiltext{2}{Department of Astronomy, University of Toronto, 60 St George Street, Toronto M5S 3H8, Canada}
\altaffiltext{3}{CFHT Legacy Survey Postdoctral Fellow}
\altaffiltext{4}{Steward Observatory, University of Arizona, 933 North Cherry Avenue, Tucson, AZ 85721.}
\altaffiltext{5}{Physics Department, University of California, Davis, CA 95616.}
\altaffiltext{6}{IGPP-LLNL, L-413, 7000 East Avenue, Livermore, CA 94550.}
\altaffiltext{7}{Institute of Astronomy, University of Tokyo, 2-21-1 Osawa, Mitaka, Tokyo 181-0015, Japan.}
\altaffiltext{8}{Department of Physics, University of Tokyo, Hongo 7-3-1,
Bunkyo-ku, Tokyo 113-0033, Japan.}
\altaffiltext{9}{Princeton University Observatory, Peyton Hall, Princeton, NJ 08544.}
\altaffiltext{10}{Department of Physics \& Astronomy, York University, 4700 Keele St., Toronto, ON, M3J 1P3, Canada}
\altaffiltext{11}{Department of Astronomy and Astrophysics, The Pennsylvania State University, University Park, PA 16802}  
\altaffiltext{12}{Apache Point Observatory, 2001 Apache Point Road, P.O. Box 59, Sunspot, NM 88349-0059}
\altaffiltext{13}{Mt. Suhora Observatory, Cracow Pedagogical University, ul. Podchorazych 2,30-084 Cracow, Poland}
\altaffiltext{14}{Fermi National Accelerator Laboratory, P.O. Box 500, Batavia, IL 60510.}
\altaffiltext{15}{Department of Astronomy and Astrophysics, The University of Chicago, 5640 South Ellis Avenue, Chicago, IL 60637.}

\begin{abstract}

We report follow-up observations of two gravitational lens candidates identified in the Sloan Digital Sky Survey (SDSS) dataset. We have confirmed that SDSS~J102111.02+491330.4 is a previously unknown gravitationally lensed quasar. This lens system exhibits two images of a $z = 1.72$ quasar, with an image separation of $1{\farcs}14 \pm 0.04$. Optical and near-IR imaging of the system reveals the presence of the lensing galaxy between the two quasar images. Observations of SDSS~J112012.12+671116.0 indicate that it is more likely a binary quasar than a gravitational lens. This system has two quasars at a redshift of $z = 1.49$, with an angular separation of $1{\farcs}49 \pm 0.02$. However, the two quasars have markedly different SEDs and no lens galaxy is apparent in optical and near-IR images of this system. We also present a list of 31 SDSS lens candidates which follow-up observations have confirmed are \textit{not} gravitational lenses.  

\end{abstract}

\section{Introduction}

The fortuitous alignment of the gravitational potential of a foreground galaxy with a background quasar can cause multiple images of the quasar to be seen. These rare systems can be used to probe to mass distribution of the lens galaxy \citep{ 1995ApJ...445..559K}, measure the Hubble constant through time delay measurements \citep{1964MNRAS.128..307R,1997ApJ...482...75K,1999ApJ...527..513K}, and constrain cosmological models through the use of lensing statistics \citep{1990ApJ...365L..43T, 1990MNRAS.246P..24F,2002PhRvL..89o1301C,2004ApJ...601..104K}. For a comprehensive review of the principles and applications of strong lensing, the reader is referred to \citet{2004kochanek_saasfee}. Newly discovered lens systems can be of interest both for individual study and as additional datapoints in studies of lens ensembles. In some cases, lens samples can be used to draw statistical inferences without particular regard to how the lenses were selected (eg \nocite{2005ApJ...623..666R,2004ApJ...611..739T}Treu \& Koopmans 2004, Rusin \& Kochanek 2005), but studies which consider properties of the lens population, such as the distribution of image separations (eg \nocite{2002ApJ...566..652L,2003ApJ...599L..61C} Li \& Ostriker 2002, Chae and Mao 2003), or the quad fraction \citep{2004ApJ...608...25C}, would benefit from a sample defined by a clearly understood selection procedure, so as to allow for the correction of selection effects. To date, the best-defined sample is that produced by the CLASS radio survey \citep{2003MNRAS.341....1M}. However, the fact that only $\sim 10\%$ of quasars are radio-loud \citep{2002AJ....124.2364I} implies that optical surveys offer the best future prospect for assembling large statistical samples.  

The Sloan Digital Sky Survey \citep[henceforth SDSS]{2000AJ....120.1579Y} dataset provides an excellent opportunity to discover gravitationally lensed quasars. The $\sim 10^4$ deg$^2$ of five-band photometry and the spectroscopic sample of $\sim 10^5$ quasars should allow for the compilation of a lensed quasar sample which is both large and statistically well-defined in comparison to existing optical samples. The parent sample of SDSS quasars is selected by a single clearly defined algorithm \citep{2002AJ....123.2945R} and the uniform image processing pipeline allows for geometric selection effects to be accurately characterized through the use of image simulations (Pindor et al. 2003, henceforth P03\nocite{2003AJ....125.2325P}). To date, the SDSS has discovered several galaxy-scale lenses (\nocite{2003AJ....126.2281J,2004AJ....127.1318P}eg. Johnston et al. 2003, Pindor et al. 2004) as well as the largest known quasar lens \citep{2003Natur.426..810I}. In this work, we report follow-up observations of two objects identified as promising lens candidates during an MMT spectroscopic run. We also present a list of objects which have been ruled out as gravitational lenses during our ongoing search for lensed quasars in the SDSS dataset.

\section{Observations}

\subsection{SDSS Technical Summary}

The SDSS is a photometric and spectroscopic survey across 10,000 square degrees of the northern Galactic cap using the 2.5m SDSS telescope at Apache Point Observatory. SDSS imaging is carried out with a wide-field camera \citep{1998AJ....116.3040G} which makes nearly simultaneous observations of objects in five passbands: {$u$ $g$ $r$ $i$ $z$}. Together, the passbands cover the optical wavelengths from the atmospheric cut-off in the blue to the minimum detectable energy for the silicon CCDs in the red \citep{1996AJ....111.1748F}. Photometric calibration of the imaging survey is separately carried out by an automated 0.5m telescope which monitors a set of standard stars \citep{2002AJ....123.2121S} while photometric data is being acquired \citep{2001AJ....122.2129H}. The position of the stellar locus can be used to independently verify the accuracy of the photometric zeropoint calibrations \citep{2004AN....325..583I}. The SDSS imaging camera also incorporates astrometric CCDs which provide astrometry of detected objects with an accuracy typically better than 0$\farcs1$ \citep{2003AJ....125.1559P}. SDSS spectroscopy is carried out on the same telescope by two fiber-fed double spectrographs which produce spectra with a resolution ($\lambda / \Delta \lambda$) of $\sim$ 2000 covering the wavelength range 3800--9200 \AA. Together, the spectrographs have 640 fibers which are assigned based on previous SDSS imaging through an efficient tiling algorithm \citep{2003AJ....125.2276B}. For more comprehensive documentation of the survey, readers are encouraged to consult \citet{2002AJ....123..485S}, \citet{2003AJ....126.2081A}, \citet{2004AJ....128..502A}, and \citet{2005AJ....129.1755A}.

\subsection{Selection as Lens Candidates} 

Five-band photometry of SDSSJ~102111.02+491330.4 (henceforth SDSS~J1021+4913) and SDSS~J112012.12+671116.0 (henceforth SDSS~J1120+6711), respectively, was obtained in the course of normal SDSS imaging on 2001 December 19\footnote{The SDSS photometric designation of SDSS~J1021+4913 is (run/rerun/camCol/field/id) 2830/41/5/294/52} and 2000 April 27\footnote{The SDSS photometric designation of SDSS~J1120+6711 is (run/rerun/camCol/field/id) 1412/40/4/131/6}. On the basis of their broad-band colors, both objects were identified as quasar candidates by SDSS quasar target selection. SDSS spectroscopy on 2002 March 13 confirmed SDSS~J1021+4913 to be a quasar at a redshift of $z = 1.72$\footnote{The SDSS spectroscopic designation of SDSS~J1021+4913 is (mjd/plate/fibre) 52347/873/560}, and on 2001 February 02 confirmed SDSS~J1120+6711 to be a quasar at a redshift of $z = 1.49$\footnote{The SDSS spectroscopic designation of SDSS~J1120+6711 is (mjd/plate/fibre) 51942/491/431}. 

Having been confirmed as quasars, both objects were selected as lens candidates by the selection algorithm described by P03. This algorithm uses the point spread function (PSF), as measured by the SDSS photometric pipeline \citep{2001adass..10..269L}, to fit the postage stamp image of each $z > 0.6$ SDSS quasar to both a one and two point source model. Lens candidates, which are expected to exhibit multiple images, are required to show a significant improvement in goodness-of-fit when modelled with two point sources, as compared to one. Further, selected candidates are required to have a best-fit component separation of $< 3^{\prime\prime}$, to have a consistent geometry when fitted independently in $g$,$r$, and $i$, and to exhibit a color difference between quasar images not exceeding that allowed by a simple reddening model. Our two PSF photometric models of the SDSS atlas images predicted for SDSS~J1021+4913 a component separation of $1\farcs07$ and a flux ratio of A/B $\approx$ 2:1, and for SDSS~J1120+6711 a component separation of $1\farcs75$ and a flux ratio of A/B $\approx$ 4:1. We further note that both candidates were also independently identified by the selection algorithm of \citet{Inada1}. 

\subsection{Spectroscopic Observations}

We observed both objects on 2003 April 6, using the 6.5m Multiple Mirror Telescope (MMT). Spectra were obtained using the blue channel of the MMT spectrograph with the 300/mm grating and $1^{\prime\prime}$ slit, yielding a spectral resolution of 6 \AA. The slit was aligned along the separation axis of the pairs, as predicted from SDSS imaging by a photometric model consisting of two point sources (see P03), so that the two components were observed simultaneously. The integration time was 600s for SDSS~J1021+4913 and 900s for SDSS~J1120+6711. 

Although the seeing at the time of observation was approximately $1^{\prime\prime}$, the small separation between the quasar pairs led to their only being marginally resolved in the focal plane. Consequently, we implemented a deblending procedure to minimize cross-contamination between the extracted 1D spectra. The spectra were deblended as follows:

- The 2D spectrum was binned into 20 bins along the dispersion axis and the flux in each of these bins was integrated to produce a series of 20 spatial cross-sections.

- A profile consisting of two Gaussians was fitted to each of the spatial cross-sections, subject to the constraint that the separation between the two Gaussians be the same for each bin. 

- Apertures for the two components were defined by assigning a fraction of the flux from each pixel to the first (second) aperture corresponding to the fractional contribution of the first (second) Gaussian to the model profile at that pixel. Hence, the model profiles can be thought of as relative weights which determine how much of the flux in a given pixel should be assigned to either of the two apertures.

This procedure is designed to ensure that the extracted apertures accurately trace the positions of the two objects on the 2D image. However, for objects such as SDSS~J1021+4913, where the images of the two components overlap significantly in the focal plane, it cannot all together eliminate cross-contamination. Fortunately, we can estimate and correct for the presence of contaminating flux as follows:

- Two additional Gaussian models (sets of weights) are constructed, each of which is a spatial reflection of the one of the original pair of Gaussian models with respect to the position of the other. For instance, the first reflected model is identical in shape to the first real model but appears on the other (spatial) side of the the second real model. 

- Two reflected apertures are then assigned flux as above, using one reflected and one real set of weights for each. For instance, flux is assigned to the first reflected aperture using the first reflected and second real Gaussians.  

\begin{figure}
\plotone{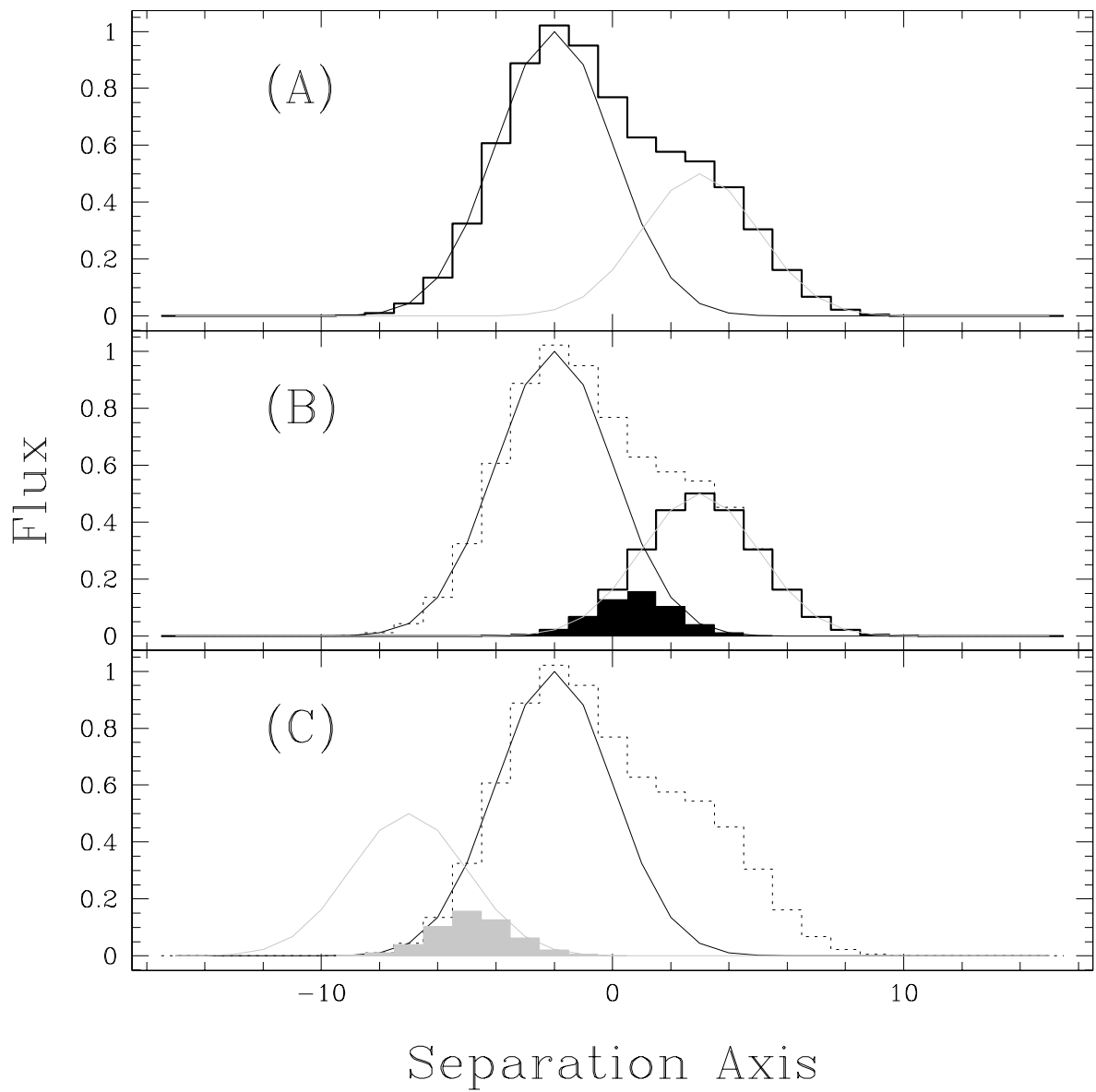}
\caption{(A): The combined flux (solid histogram) of two marginally resolved sources along the separation axis. The two Gaussian curves represent the primary and secondary fitted weights. (B): As in (A), expect that the combined flux is indicated by the dotted histogram. Also shown is the flux assigned to the secondary aperture (solid unfilled histogram) according to the two weights, and the fraction of this assigned flux (filled histogram) which originates from the primary object. (C): The weight corresponding to the secondary object has been reflected onto the other side of the primary object, and flux has been assigned to a reflected aperture using this reflected and the original primary weight. The assigned flux (shaded histogram) is equal to, and can be used to correct for, the contaminating flux shown in (B).}
\label{ref_ap}
\end{figure}

Assuming that the PSF is essentially symmetric along the separation axis, these reflected apertures now contain the same amount of contaminating flux from the adjacent object as the real (unreflected) apertures. Cross-contamination can be corrected for by subtracting the flux in each reflected aperture from that in the corresponding real aperture. Figure \ref{ref_ap} schematically illustrates our deblending procedure. 

\begin{figure}
\plotone{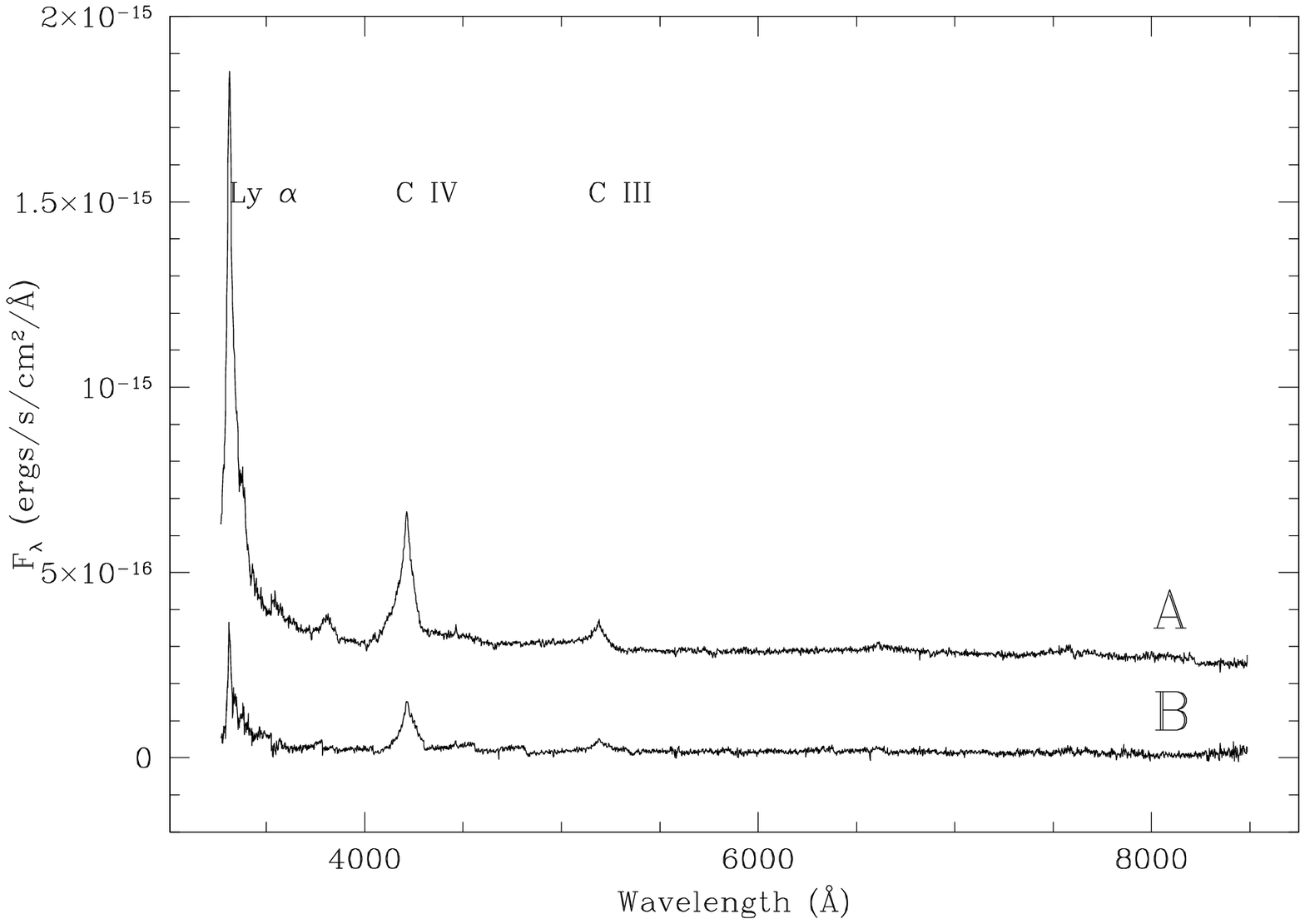}
\caption{MMT spectra of the two components of SDSS~J1021+4913. The spectrum of the brighter component has been shifted upward by an arbitrary offset to facilitate comparison.}
\label{1021_spec}
\end{figure}

Figure \ref{1021_spec} shows the spectra of the two components of SDSS~J1021+4913. We cross-correlated the spectra and estimated the velocity difference between them be less than the dispersion per pixel, corresponding to $ \sim 100$ km~s$^{-1}$, and consistent with zero. The most significant spectral difference between the two components is that component B appears considerably redder at the blue end than component A. This qualitative assessment is supported by measurements of the equivalent widths of emission lines in the two components (Primary: Ly $\alpha$ (450$\pm$100~\AA) \ion{C}{4} (400$\pm$10~\AA) \ion{C}{3} (105$\pm$10~\AA) Secondary: Ly $\alpha$ (300$\pm$50~\AA) \ion{C}{4} (390$\pm$30~\AA) \ion{C}{3} (95$\pm$10~\AA)), although there are large uncertainties associated with determining an appropriate continuum level for Ly $\alpha$. This discrepancy can be explained as being due to varying extinction along different sightlines through the lens galaxy. Note that the feature at 7600 \AA\ is atmospheric. 

\begin{figure}
\plotone{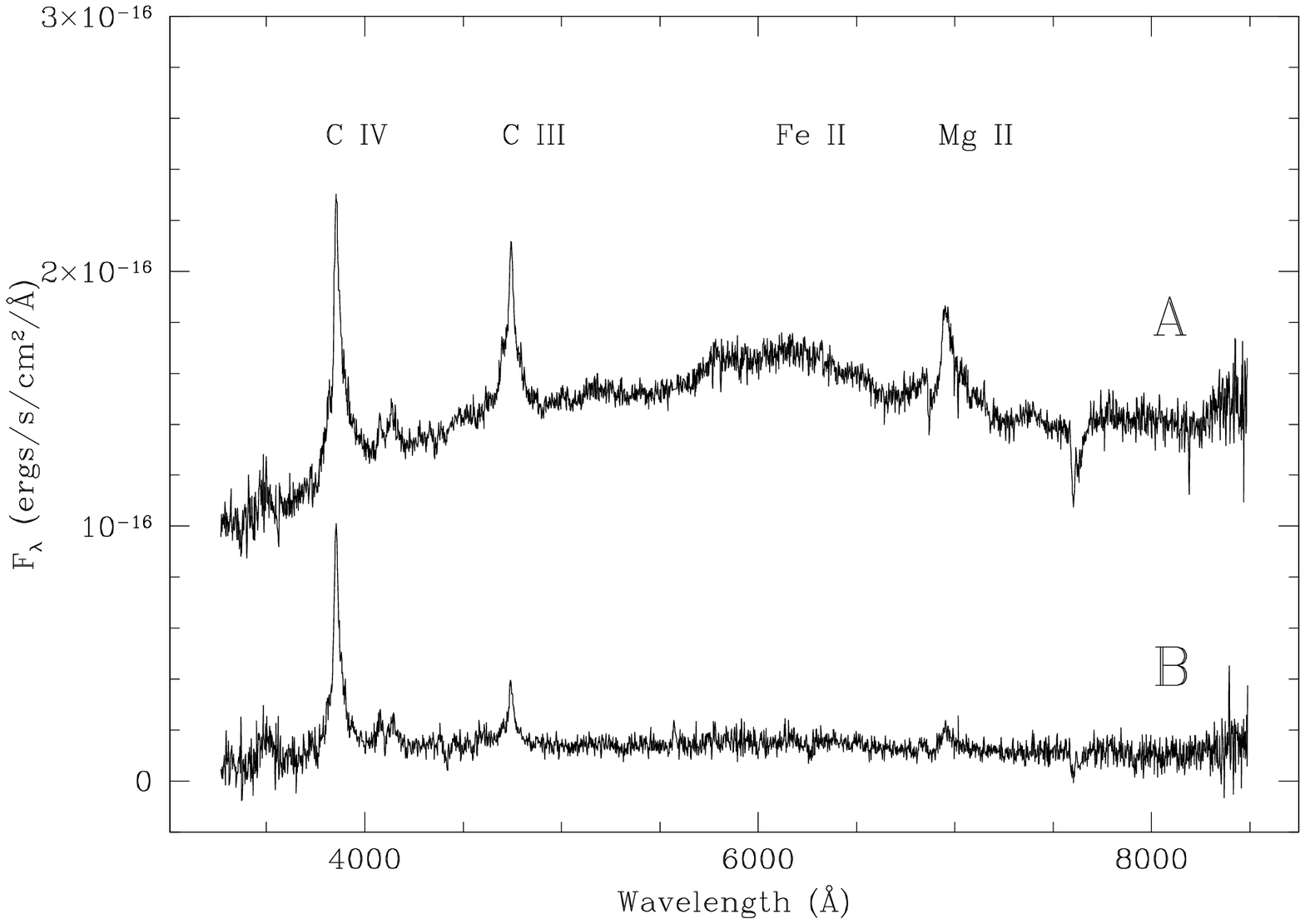}
\caption{MMT spectra of the two components of SDSS~J1120+6711. The spectrum of the brighter component has been shifted upward by an arbitrary offset to facilitate comparison.}
\label{1120_spec}
\end{figure}

Figure \ref{1120_spec} shows the spectra of the two components of SDSS~J1120+6711. We cross-correlated the spectra and also estimated the velocity difference between them to be less than 100 km~s$^{-1}$. The spectra of the two components differ significantly; the continuum of the brighter component is much redder at the blue end, the strength of the \ion{C}{4} line relative to the other lines is much greater in the fainter component, and there is no evidence in the fainter component of the broad \ion{Fe}{2} complex which is seen blueward of \ion{Mg}{2} in the brighter component.  

In summary, spatially-resolved spectroscopy revealed that both SDSS~J1021+4913 and SDSS~J1120+6711 are quasar pairs with indistinguishable redshifts. However, the two components of SDSS~J1021+4913 have similar spectral energy distributions (SEDs), consistent with the lensing hypothesis, while the two components of SDSS~J1120+6711 have significantly different SEDs.   

\subsection{Near-IR Imaging}   

We subsequently obtained near-IR imaging of these two systems in order to search for the existence of a possible lens galaxy in each case. 

\subsubsection{SDSS~J1021+4913}

 On 2003 August 12, we imaged the system in the $K^{\prime}$-band using the Near Infrared Camera (NIRC, Matthews \& Soifer 1994\nocite{1994ExA.....3...77M}) on the Keck I telescope. The observations consisted of a five-point dither pattern, integrating for 30s at each pointing. The data were flattened, sky subtracted, shifted, and stacked using the DIMSUM package in IRAF.

\begin{figure}
\plotone{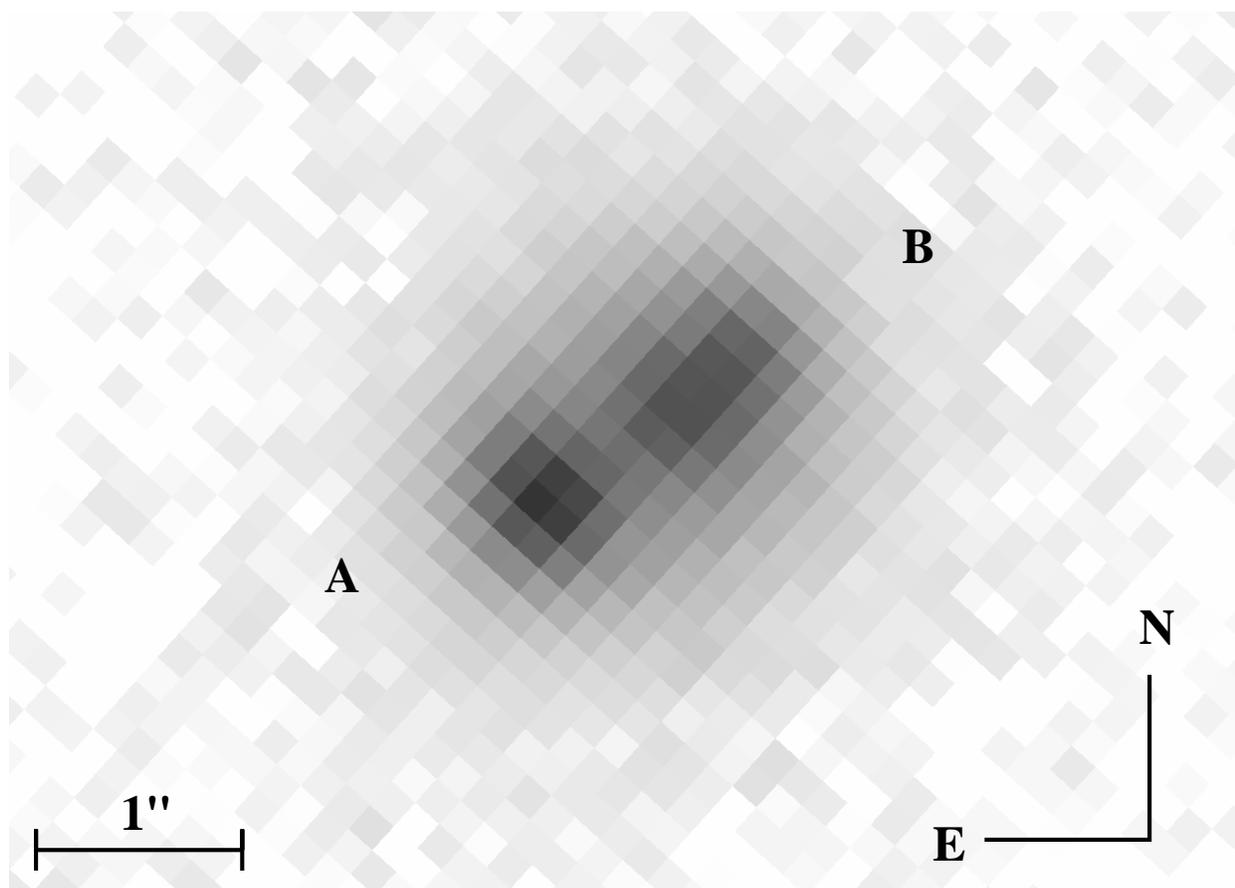}
\caption{NIRC $K^{\prime}$-band image of SDSS~J1021+4913. }
\label{1021_NIRC}
\end{figure}

Figure \ref{1021_NIRC} shows the $K^{\prime}$-band image of SDSS~J1021+4913. It is clear that the fainter component is extended relative to the brighter component. It is unlikely that this extended flux is associated with the quasar host galaxy since i) it would put the quasar at a very unusual off-center position relative to the host galaxy center-of-light, and ii) it would be an additional co-incidence that the host galaxy emission happens to concentrate between the two quasars. Instead, we interpret this extended flux as indicating the presence of a lens galaxy. The lens galaxy is bright enough that it makes impractical a simple PSF subtraction. Hence, we modelled the system as a combination of three components; two point sources, using the profile recommended by \citet{1996PASP..108..699R}, and an extended component represented by a deVaucouleurs profile of the form: 

\begin{eqnarray}
I(x,y) = I_0 \mathrm{ exp }(-7.67((x^2 + y^2)^{1/2}/R_{\rm eff})^{1/4})
\end{eqnarray}
convolved with the PSF. In general, a deVaucouleurs profile can also account for the axis ratio and position angle of a galaxy on the sky, but we chose this circularly symmetric form for simplicity. The parameters for the best-fit model are listed in Table \ref{1021_table}. The uncertainties in this model are dominated by systematic errors associated with the imperfect PSF model. The errors listed are the observed variation in the model parameters when the best-fit FWHM is varied by $\pm$ 10\%. The reduced chi-square of the best-fit two point sources plus deVaucouleurs model is 1.6 (1590 degrees of freedom). For comparison, the reduced chi-square of the best-fit two points sources (only) model is 5.5 (1594 d.o.f.). We interpret this improvement in the fit as definitive evidence for the presence of a lens galaxy, and consequently, in support of the lensing hypothesis. There are no sources in the field bright enough to allow for direct calibration to 2MASS photometry \citep{2003yCat.2246....0C}, but we can use the results of Covey et al. (2005, in preparation), who characterized the stellar locus in optical and near-IR colors, to predict the $K'$-band magnitudes of SDSS-detected stars in the field. The errors in the derived magnitudes listed in Table \ref{1021_table} include the width of the stellar locus in the relevant optical-IR color space. However, we do not have enough data to characterize possible systematic errors in this photometric calibration, which may be significant compared to the random errors quoted.

The best-fit photometric model indicates that the angular separation between quasar image A and G(alaxy) is nearly the same as BG. A simple singular isothermal sphere mass model, naturally, predicts that the quasar images should have nearly equal fluxes in this case. Instead, the observed flux ratio is A/B $\approx$ 5:2. This discrepancy can be explained by i) extinction of image B relative to image A, consistent with the differential reddening noted above, or ii) modelling the lens galaxy with a more complicated mass distribution. 

To explore the latter possibility, we modelled the lens system
using two mass models; a Singular Isothermal Sphere with external shear
(SIS+shear) model and a Singular Isothermal Ellipsoid (SIE) model.
Each model has eight parameters: the Einstein radius $R_{\rm E}$, 
shear or ellipticity ($\gamma$ or $e$) and its position angle
($\theta_\gamma$ or $\theta_e$), the position of the lens galaxy, 
the position of the source quasar, and the flux of the source quasar.
The number of observables is also eight, therefore the number of degrees 
of freedom is zero. We fit the mass models with the public software 
{\sl lensmodel} \citep{keeton01}. As expected, both models yield almost 
perfect fits, chi-square of $\sim 0$. The resulting best-fit values are 
summarized in
Table \ref{massmodel_table}. We find that we can reproduce the photometric 
observations with mass models which have realistically small shear/ellipticity,
 $\gamma=0.01$ or $e=0.05$. Thus the lensing
hypothesis is also supported by our mass modelling of the system. We
estimated the expected time delays between images and find that they are
quite small, $\sim 1h^{-1}{\rm day}$ (B leads A), assuming a lens
redshift of $z=0.5$. This is because of almost the equal angular
separation between AG and BG.  

\subsubsection{SDSS~J1120+6711}

On 2004 May 8, we imaged SDSS~J1120+6711 in the $H$-band at the MMT using 
the ARIES instrument \citep{1998SPIE.3354..750M} and the f/15 MMT/AO adaptive secondary \citep{2002bcao.conf...55R}. There is 
no guide star close enough to permit closed-loop imaging, but we 
used a bright star $12'$ away to phase up the optical system.  We integrated for 16 minutes before clouds rolled in.  The resulting stack of images has 
a $0.4''$ FWHM.

\begin{figure}
\plotone{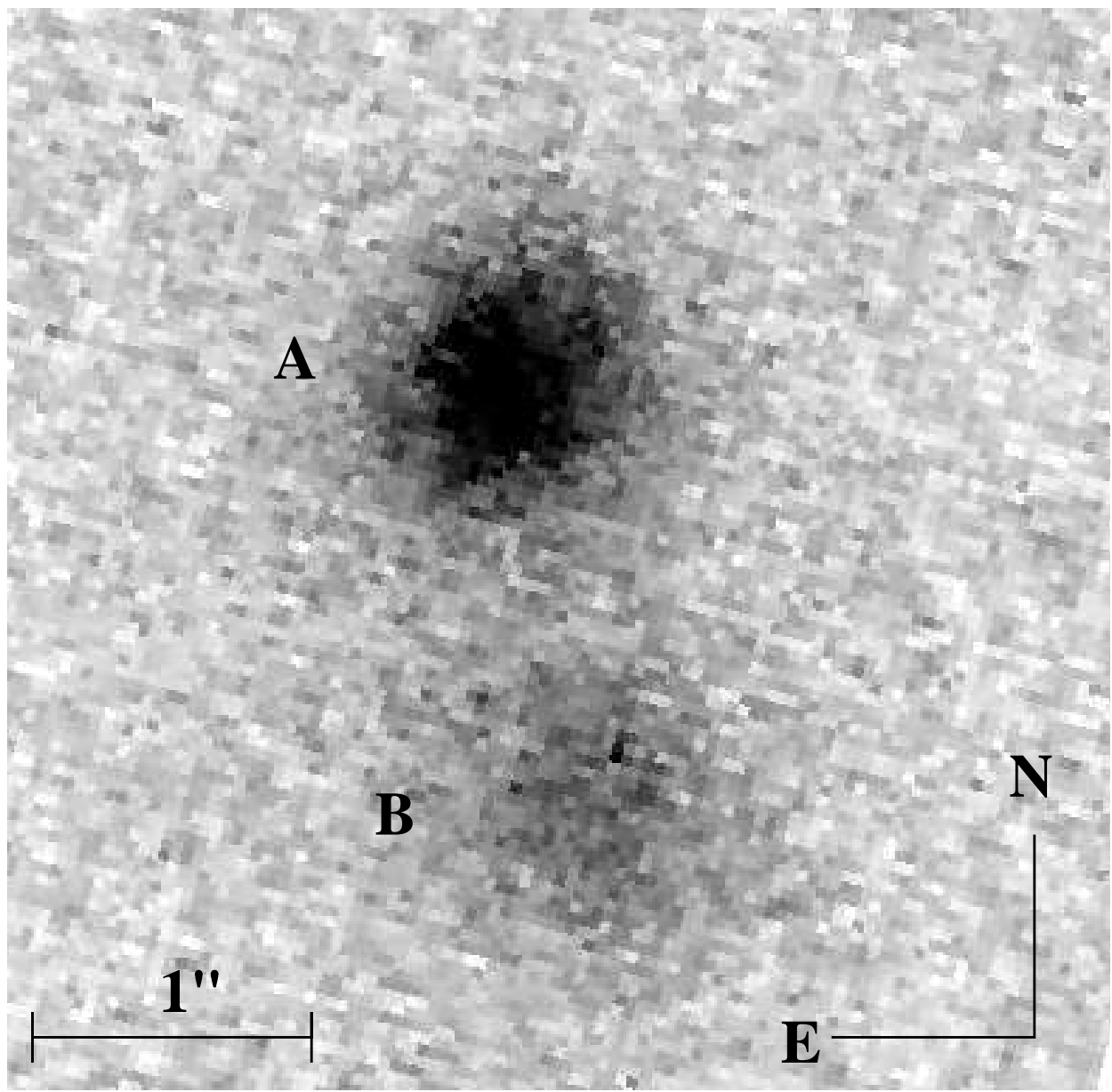}
\caption{ARIES $H$-band image of SDSS~J1120+6711.}
\label{1120_image}
\end{figure}

\begin{figure}
\plotone{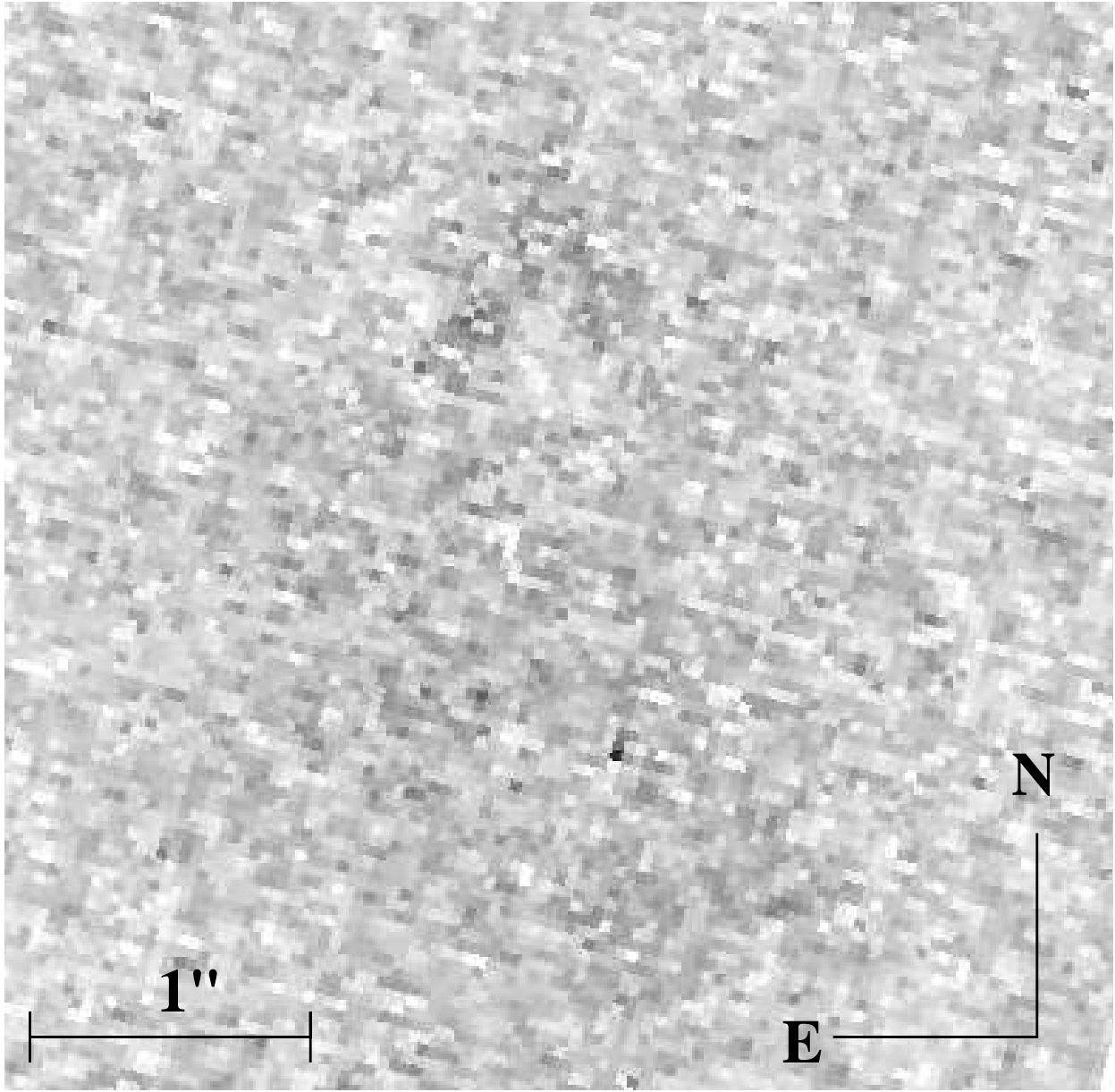}
\caption{$H$-band residuals for SDSS~J1120+6711 following subtraction of two point sources. Greyscale is the same as in Figure \ref{1120_image}.}
\label{1120_residuals}
\end{figure}

Figure \ref{1120_image} shows the H-band image of SDSS~J1120+6711. The point sources appear extended because they are highly oversampled by the minute pixel scale (0$\farcs04$), but there is no obvious indication of a lens galaxy. Indeed, a two point source model, again using the model profile of Racine (1996), fits the image very well; yielding a reduced chi-square of 1.2 (6394 d.o.f.) and leaving residuals, shown in Figure \ref{1120_residuals}, with no obvious flux concentrations. The parameters for this photometric model are listed in Table \ref{1120_table}. Again, the magnitudes listed are derived by estimating the optical-infrared colors of stars in the field from their SDSS colors.  

Having subtracted the point sources, we set an upper limit on the flux of any remaining sources by introducing simulated point sources into the residual image and visually inspecting these simulated residuals. We thus estimated that any remaining sources could have a point source flux of not more than 10\% of the combined flux of components A and B. This upper limit corresponds to a magnitude of $H > 19.4 \pm 0.1$.

We can compare this upper limit to the results of \citet{2003ApJ...587..143R} who found that the observed magnitudes of lens galaxies can be predicted from their observed image separations through the relation 

\begin{eqnarray}
m_{obs} = M_{*0} + DM + 2.5 \gamma_{E+K}z_d - 1.25\gamma_{FJ}\mathrm{log}\Delta\theta_{red}
\end{eqnarray}

where $DM$ is the distance modulus, $\gamma_{E+K}$ is a parameter subsuming evolution and spectral K-corrections, $\gamma_{FJ}$ is the slope of the Faber-Jackson relation in the chosen band, $z_d$ is the lens galaxy (deflector) redshift, and $\Delta\theta_{red}$ is the reduced image separation, $\Delta\theta_{red} \equiv (\Delta\theta / \Delta\theta_{*})(D_s/D_{ds})$, where $\Delta\theta_{*}$ is the image separation produced by a singular isothermal sphere with a velocity dispersion of 225 km~s$^{-1}$, and $D_s$ and $D_{ds}$ are the angular diameter distances between the observer and source and deflector and source, respectively. The $H$-band values of $M_{*0} (=-23.8 + 5 \mathrm{log}h)$, $\gamma_{E+K} (=-0.21)$, and $\gamma_{FJ} (=3.32)$ are all given by Rusin et al., based upon HST photometry and spectrophotometric modelling of 28 early-type lens galaxies. The observed rms scatter is $\sim 0.5$ magnitudes. This relation predicts that given an observed source redshift and image separation, there exists a redshift at which the lens galaxy has a maximum apparent magnitude. In the case of SDSS~J1120+6711, the maximum occurs at $z \sim 0.75$ and implies that $H < 18$. In short, we did not detect any indication of a lens galaxy in our $H$-band image of SDSS~J1120+6711, and a simple estimate indicates that this non-detection is inconsistent with the presence of a lens galaxy of typical luminosity.

\subsection{Optical Imaging} 

We further obtained deep $I$-band images of these two objects using the 8k mosaic CCD on the University of Hawaii 2.2m telescope (UH88) at Mauna Kea. The imaging was conducted on 2004 May 23 (SDSS~J1021+4913) and 25 (SDSS~J1120+6711). The integration times were 120 sec for SDSS~J1120+6711 and 360 sec for SDSS~J1021+4913. We estimate the FWHM of these images to be $0\farcs8$. Note that the optical and near-IR images were independently analyzed by two of the authors, Inada and Pindor respectively, and discrepancies between the photometric models are most likely indicative of the systematic errors present. 

\subsubsection{SDSS~J1021+4913}

\begin{figure}
\plotone{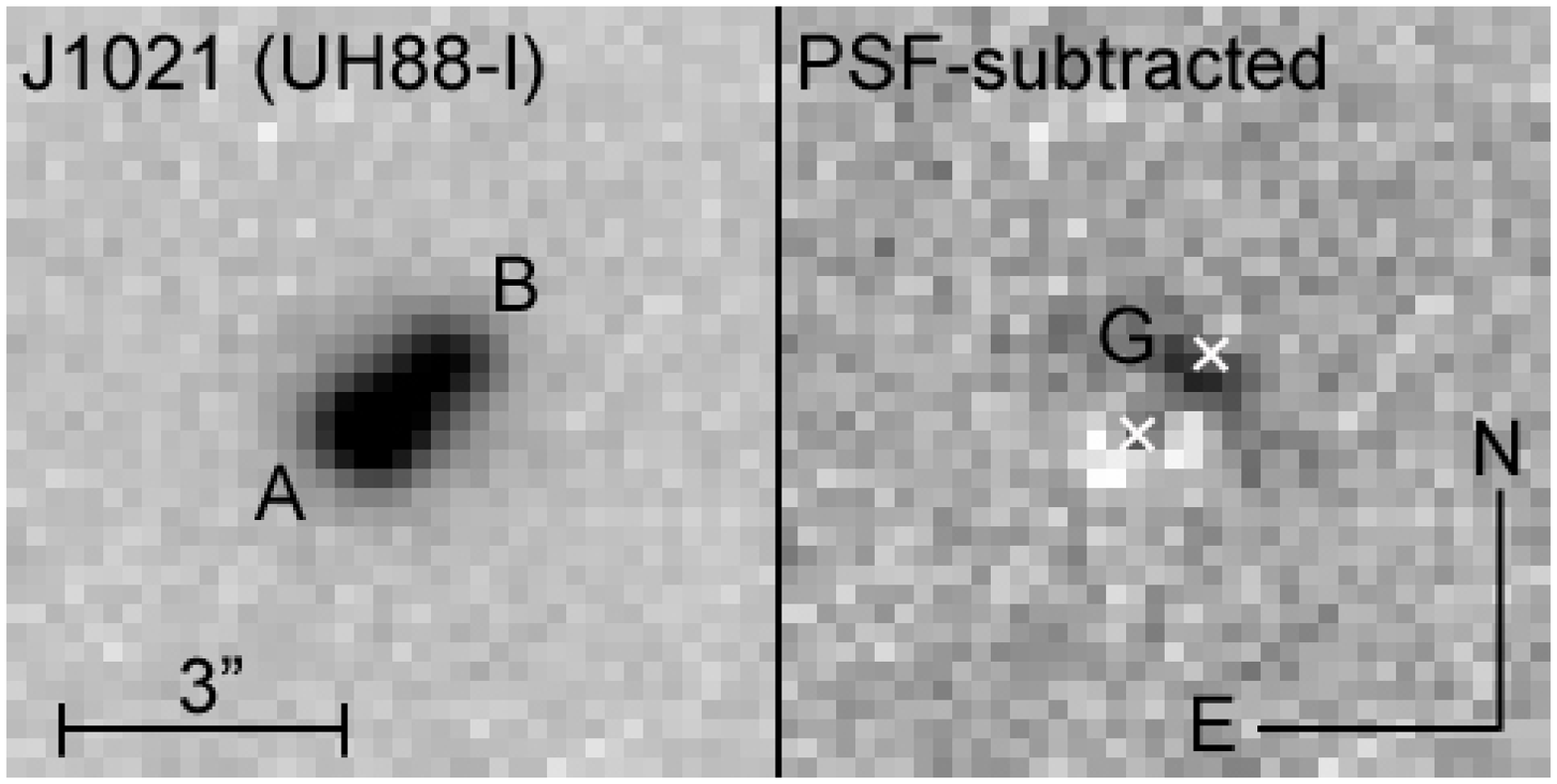}
\caption{UH88 $I$-band image of SDSS~J1021+4913. The white crosses in the PSF-subtracted image indicate the location of the point source components.}\label{1021_uh88}
\end{figure}

Figure \ref{1021_uh88} shows the $I$-band image of SDSS~J1021+4913, both as observed and following the subtraction of two point sources. For the optical data, the point source subtraction was performed using a Gaussian PSF model whose FWHM was determined by fitting to nearby stars. Table \ref{1021_table} lists the corresponding photometric model. Photometric calibration was obtained relative to the standard star PG 1528+062 \citep{1992AJ....104..340L}. The listed magnitude errors do not include the photometric uncertainty of the standard star. We take the close agreement between our optical and near-IR models as definitive confirmation of the presence of the lens galaxy. Certainly, the derived galaxy magnitude makes this object too bright to be a $z=1.72$ host galaxy \citep{2005astro.ph..3102D}. We can invert the relation of Rusin et al. to estimate the lens galaxy redshift, but the derived constraint $0.45 < z < 1.25$ (66\% confidence) is not very stringent.  

\subsubsection{SDSS~J1120+6711} 

\begin{figure}
\plotone{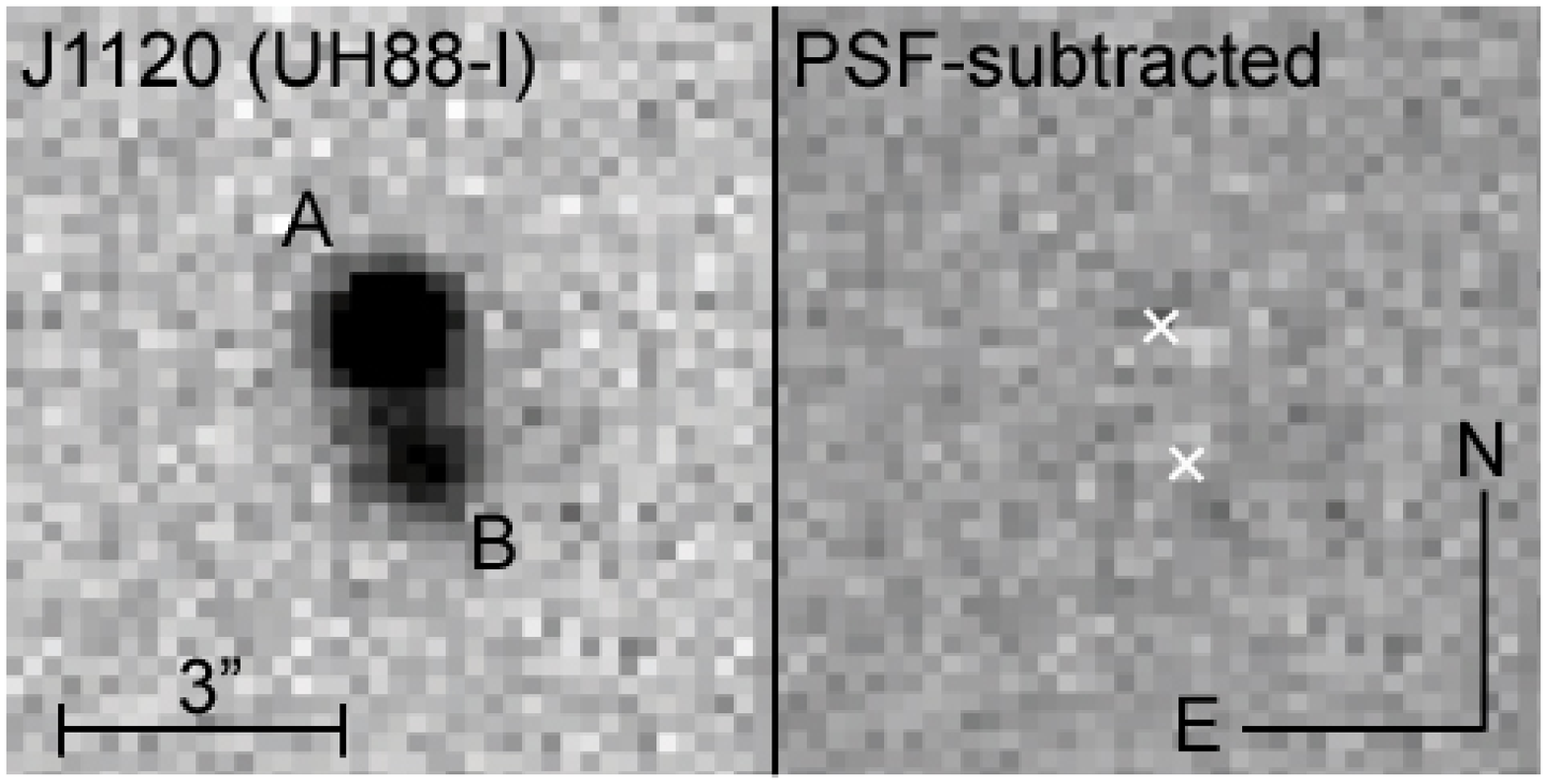}
\caption{UH88 $I$-band image of SDSS~J1120+6711. The white crosses in the PSF-subtracted image indicate the location of the point source components.}\label{1120_uh88}
\end{figure}

Figure \ref{1120_uh88} shows the $I$-band image of SDSS~J1120+6711, both as observed and following the subtraction of two point sources. Table \ref{1120_table} lists the corresponding photometric model. We are unable to detect any object in the residual image corresponding to more than a $3\sigma$ fluctuation above the sky level. We estimate that this corresponds to an upper limit of $I > 24$. Again employing the method of Rusin et al., we predict that the $I$-band magnitude of the lens galaxy should be $I < 20.5$~. In agreement with our $H$-band image, the $I$-band image of SDSS~J1120+6711 shows no evidence of a lens galaxy and is inconsistent with the presence of a lens galaxy of typical luminosity.  

\section{Conclusions}

The observations reported above have confirmed that SDSS~J102111.02+491330.4 is almost certainly a previously unknown gravitationally lensed quasar. High spatial resolution spectroscopy revealed the existence of two quasar images with an angular separation of $1{\farcs}14 \pm 0.04$ and having identical redshifts of $z = 1.72$. High resolution $K^{\prime}$-band imaging indicates the presence of a lens galaxy between the two quasar images. SDSS~J1021+4913 does not appear to have any peculiarities which would obviously compel individual studies of the system. Conversely, the fact that simple lens models readily reproduce the lens geometry implies that the system is well-suited to various ensemble studies. The utility of this system in such studies would be aided an improved photometric characterization and optical redshift measurement of the lens galaxy, both of which should be plausible for an 8m class telescope. Our mass models predict a short time delay ($\sim 1$ day), but the relatively small separation of the two quasar images makes the system less than ideal for a monitoring campaign. SDSS~J1021+4913 represents an incremental increase in what will likely soon be the largest uniformly selected sample of lensed quasars. The SDSS lens sample will have an additional merit relative to radio surveys like CLASS in that both the redshifts of the lensed sources are known and the redshift distribution of the background quasar population is well-characterized \citep{2005MNRAS.360..839R}.    

Observations of SDSS~J112012.12+671116.0 indicate that it is more likely a binary quasar than a lensed pair. Although the two components are both quasars with indistinguishable redshifts of $z = 1.49$, they have markedly different SEDs. Optical and $H$-band imaging of the system did not reveal any indication of a lens galaxy. Binary quasars can be considered a contaminant in a search for gravitational lenses, but they are interesting objects in their own right. The SDSS has previously discovered a number of other binary quasars, although mostly, due to fibre collisions, at considerably greater angular separations \citep{2003AJ....126.2579S}. A sample of SDSS selected binary quasars will provide insights on quasar clustering and the onset of the quasar phenomenon (Hennawi et al, in preparation). It has been known for some time that extrapolating observed large scale quasar-quasar clustering to small separations considerably underpredicts the number of arcsecond scale binaries \citep{1991ASPC...21..349D}. Hence, the confirmation of SDSS~J1120+6711 and other uniformly selected binaries is the best empirical method for determining to binary to lens ratio at these separations. Knowledge of this ratio is important for trying to predict the efficiency of future lens searches \citep{2005ApJ...626..649P}.

B.P.\ is supported by the Natural Sciences and Engineering Research Council of Canada. D.J.E.\ is supported by National Science Foundation grants AST-0098577 and AST-0407200, and by an Alfred P. Sloan Research Fellowship. 

Funding for the creation and distribution of the SDSS Archive has been provided by the Alfred P. Sloan Foundation, the Participating Institutions, the National Aeronautics and Space Administration, the National Science Foundation, the U.S. Department of Energy, the Japanese Monbukagakusho, and the Max Planck Society. The SDSS Web site is http://www.sdss.org/.

The SDSS is managed by the Astrophysical Research Consortium (ARC) for the Participating Institutions. The Participating Institutions are The University of Chicago, Fermilab, the Institute for Advanced Study, the Japan Participation Group, The Johns Hopkins University, the Korean Scientist Group, Los Alamos National Laboratory, the Max-Planck-Institute for Astronomy (MPIA), the Max-Planck-Institute for Astrophysics (MPA), New Mexico State University, University of Pittsburgh, University of Portsmouth, Princeton University, the United States Naval Observatory, and the University of Washington.

A portion of the results reported here were obtained at the W. M. Keck Observatory, which is operated as a scientific partnership among the California Institute of Technology, the University of California, and the National Aeronautics and Space Administration. The observatory was made possible by the generous financial support of the W. M. Keck Foundation.

We thank Craig Kulesa, Dylan Curley, and Don McCarthy for providing the ARIES instrument and supporting our infrared observations at the MMT.

%\bibliography{../../../papers/selection}

\appendix

\section{Rejected Lens Candidates}

The program of follow-up observations which confirmed the lensing nature of SDSS~J1021+4913 and several previous SDSS-identified gravitational lenses also served to reject the lensing hypothesis for numerous objects. Table \ref{nonlens_table} lists the positions of these rejected lens candidates. Typically, spatially-resolved spectroscopy has shown these objects to be QSO plus star superpositions. For objects which are not commented, we are unable to confidently identify the nature of the secondary object, but we can rule out the presence of broad quasar emission lines matching those of the primary object. The one binary quasar listed, SDSS~J1600+0000, has an obvious redshift difference. We have indicated those objects which satisfy the selection criteria of P03. Those objects which do not satisfy the P03 selection were either observed prior to finalization of the algorithm, selected by an altogether different algorithm, or were marginal candidates which were chosen due to observing conditions. From this list, we can surmise that quasar plus star superpositions dominate the population of false postives in this optical search, and that, compared to the $>$ 10 lenses which have been discovered or were previously known in the parent sample, the number of binaries is considerably smaller than the number of lenses at separations of $1-3^{\prime\prime}$. This is not an exhaustive list of those lens candidates which we have observed, but rather a list of those candidates which we are confident do not merit further observation in the context of lensing investigations. It should not in any way be taken as a statistical complement to the list of published SDSS lenses. Instead, our main motivation in publishing this list is to inform other investigators who might independantly select these objects as lens candidates of the results of these observations. Additional photometric and spectroscopic data for these objects are available through the SDSS SkyServer.

\clearpage

\clearpage

\begin{deluxetable}{cccc}
\tablewidth{0pc}
\tablecaption{Photometric Model Parameters for SDSS~J1021+4913.}
\tablehead{
\colhead{} & \colhead{Quasar A} & \colhead{Quasar B} & \colhead{Galaxy}
}
\startdata
\cutinhead{$H$-band : Two Analytic PSFs plus deVaucouleurs}
\textbf{Relative RA ($^{\prime\prime}$)} & 0 & -0.830 $\pm$ 0.029 & -0.270 $\pm$ 0.003 \\
\textbf{Relative Dec ($^{\prime\prime}$)} & 0 & 0.776$\pm$ 0.029 & 0.514 $\pm$ 0.014 \\
\textbf{Relative Flux\tablenotemark{a}}  & 1 & 0.38 $\pm$ 0.03 & 1.36 $\pm 0.43$\\
\textbf{Magnitude} & 18.4 $\pm$ 0.1 & 19.4 $\pm$ 0.1 & 18.0 $\pm$ 0.3 \\
\cutinhead{$I$-band : Two Analytic PSFs}
\textbf{Relative RA ($^{\prime\prime}$)} & 0 & -0.796 $\pm$ 0.023 & -0.499 $\pm$ 0.120 \\
\textbf{Relative Dec ($^{\prime\prime}$)} & 0 & 0.784$\pm$ 0.023 & 0.633 $\pm$ 0.120 \\
\textbf{Magnitude}  & 19.02 $\pm$ 0.06 & 20.01 $\pm$ 0.10 & 21.72 $\pm$ 0.50\\
\enddata
\tablenotetext{a}{Within measured effective radius}
\label{1021_table}
\end{deluxetable}

\clearpage

\begin{deluxetable}{ccccc}
\tablewidth{0pt}
\tablecaption{Mass Models for SDSS~J1021+4913\tablenotemark{a}\label{table:model}}
\tablehead{\colhead{Model} & \colhead{$R_{\rm E}$[arcsec]} &
 \colhead{$e$ or $\gamma$} &
 \colhead{$\theta_e$ or $\theta_\gamma$[deg]\tablenotemark{b}} & \colhead{$\Delta
 t$[$h^{-1}$day]\tablenotemark{c}}} 
\startdata
SIS$+$shear & $0.59$ & $0.01$ & $20.4$ & $1.2$ \\
SIE         & $0.59$ & $0.05$ & $20.1$ & $1.4$ \\
\enddata
\tablenotetext{a}{Derived from $H$-band photometry listed in Table \ref{1021_table}.}
\tablenotetext{b}{Each position angle is measured East of North.}
\tablenotetext{c}{Positive time delays mean B leads A. We assumed a
 typical lens redshift of $z=0.5$.}
\label{massmodel_table}
\end{deluxetable}

\clearpage

\begin{deluxetable}{ccc}
\tablewidth{0pc}
\tablecaption{Photometric Model Parameters for SDSS~J1120+6711.}
\tablehead{
\colhead{} & \colhead{Quasar A} & \colhead{Quasar B}}
\startdata
\cutinhead{$H$-band : Two Analytic PSFs}
\textbf{Relative RA ($^{\prime\prime}$)} & 0 & -0.30 $\pm$ 0.02 \\
\textbf{Relative Dec ($^{\prime\prime}$)} & 0 & -1.46$\pm$ 0.02 \\
\textbf{Relative Flux}  & 1 & 0.42 $\pm$ 0.01 \\
\textbf{Magnitude} & 17.3 $\pm 0.1$ & 18.3 $\pm 0.1$ \\
\cutinhead{$I$-band : Two Analytic PSFs}
\textbf{Relative RA ($^{\prime\prime}$)} & 0 & -0.353 $\pm$ 0.008 \\
\textbf{Relative Dec ($^{\prime\prime}$)} & 0 & -1.498$\pm$ 0.008 \\
\textbf{Magnitude}  & 18.49 $\pm$ 0.02 & 19.57 $\pm$ 0.04 \\
\enddata
\label{1120_table}
\end{deluxetable}

\clearpage

\begin{deluxetable}{cccl}
\tabletypesize{\scriptsize}
\tablecaption{Positions of Rejected SDSS Lens Candidates}
\tablehead{
\colhead{RA} & \colhead{Dec} & \colhead{P03} & \colhead{Comment}
}
\startdata
$08:24:04.42 $ & $+37:18:54.7 $ & Y & QSO + ELG\\
$08:45:12.73 $ & $+54:34:21.5 $ & Y & QSO + STAR?\\
$10:11:26.67 $ & $+00:43:19.2 $ & N & Ruled out by imaging\\
$10:30:21.94 $ & $+58:55:13.9 $ & Y & QSO + ELG\\
$10:41:22.84 $ & $-00:56:18.5 $ & N & QSO + STAR\\
$11:12:25.73 $ & $-00:51:01.6 $ & N & \\
$11:36:13.39 $ & $+03:38:41.1 $ & Y & QSO+STAR\\
$12:14:05.13 $ & $+01:02:05.0 $ & N & \\
$12:51:41.91 $ & $+03:11:40.9 $ & Y & QSO+STAR\\
$13:08:54.29 $ & $-02:36:13.4 $ & N & \\
$13:39:04.34 $ & $+00:22:22.1 $ & Y & \\
$14:33:16.39 $ & $+00:51:45.7 $ & Y & QSO+STAR\\
$14:34:52.73 $ & $-00:28:28.1 $ & Y & QSO+GAL?\\
$14:41:45.10 $ & $+02:37:43.3 $ & Y & QSO+STAR\\
$14:46:19.29 $ & $+00:53:17.9 $ & N &\\
$15:08:24.22 $ & $-00:06:03.9 $ & Y &\\
$15:12:36.94 $ & $+55:39:01.4 $ & Y & QSO+WD?\\
$15:15:38.95 $ & $-00:12:40.8 $ & N & \\
$15:21:19.68 $ & $-00:48:18.7 $ & N & \\
$15:25:55.81 $ & $+01:08:35.5 $ & N & QSO+STAR\\
$15:30:28.56 $ & $+51:15:14.4 $ & Y & QSO + QSO?\\
$15:41:07.46 $ & $-00:37:16.1 $ & Y & QSO+STAR\\
$16:00:15.51 $ & $+00:00:45.5 $ & Y & Binary Quasar, $z = 1.01$, $\Delta z = 0.003$, Angular Separation $1{\farcs}8$\\
$16:05:12.00 $ & $-00:47:49.6 $ & Y & QSO+STAR\\
$16:24:15.28 $ & $+00:32:51.1 $ & Y & QSO+WD?\\
$17:17:47.57 $ & $+59:32:58.2 $ & N & \\
$17:23:08.15 $ & $+52:44:55.5 $ & Y & QSO+STAR\\
$17:24:06.47 $ & $+64:07:10.7 $ & N & \\
$17:25:50.89 $ & $+56:58:59.9 $ & Y & \\
$21:29:06.24 $ & $-07:16:13.3 $ & Y & QSO+STAR\\
$23:11:16.98 $ & $-10:38:49.7 $ & Y & QSO+STAR\\
\enddata
\label{nonlens_table}
\end{deluxetable}

\end{document}